\documentclass[11pt]{article} \usepackage{a4}

\usepackage{amssymb}
\usepackage{color}
\usepackage{mathptmx,times} 

\usepackage{graphicx}
\usepackage[
  pdftitle={The ray attack on RSA},pdfauthor={Andreas de Vries},%
  pdfborder={0 0 0},%
  pdfpagemode=FirstPage,%
  dvips=true,ps2pdf=true%
]{hyperref}
  \DeclareGraphicsExtensions{.eps}

\newtheorem{satz}{Theorem}[section]
\newtheorem{lem}[satz]{Lemma}
\newtheorem{definition}[satz]{Definition}
\newtheorem{cor}[satz]{Corollary}
\newtheorem{rem}[satz]{Remark}
\newtheorem{beisp}[satz]{Example}

\newenvironment{defi}{\begin{definition}\begin{em}}{\end{em}\end{definition}}
\newenvironment{bsp}{\begin{beisp}\begin{em}}{\end{em}\end{beisp}}
\newenvironment{bem}{\begin{rem}\begin{em}}{\end{em}\end{rem}}

\begin{document}

\author{Andreas de Vries\thanks{
	e-Mail: 
	\href{mailto://de-vries@fh-swf.de}{\texttt{\scriptsize de-vries@fh-swf.de}}%
}
\\
\footnotesize{\em
FH S\"udwestfalen University of Applied Sciences,
Haldener Stra{\ss}e 182,
D-58095 Hagen}}
\title{\bf The ray attack, an inefficient trial to break RSA cryptosystems%
\footnote{This paper is a slight modification of \cite{de-Vries-2003}}
}
\date{}

\maketitle

\begin{abstract}
	The basic properties of RSA cryptosystems and some
	classical attacks on them are described. Derived from geometric
	properties of the Euler functions, the \emph{Euler function rays},
	a new ansatz to attack RSA
	cryptosystems is presented. A resulting, albeit inefficient, 
	algorithm is given. It essentially consists of a loop with
	starting value determined by
	the Euler function ray and with step width given by a function 
	$\omega_e(n)$ being a multiple of the order $\mathrm{ord}_n(e)$,
	where $e$ denotes the public key exponent and $n$ the RSA modulus.
	For $n=pq$ and an estimate $r<\sqrt{pq}$ for the smaller prime 
	factor $p$, the running time is given by 
	$T(e,n,r) = O((r-p)\ln e \ln n \ln r).$
\end{abstract}

\setcounter{tocdepth}{2}
\tableofcontents

\section{Introduction}
Since the revolutionary idea of asymmetric cryptosystems was born in
the 1970's, due to Diffie and Hellman \cite{Diffie-Hellman-1976}
and Rivest, Shamir and Adleman \cite{Rivest-et-al-1978},
public key technology became an indispensable
part of contemporary electronically based communication. 
Its applications range from authentication to digital signatures and
are widely considered to be an essential of future applications 
for e-commerce.

The most popular cryptosystem is RSA. There has been numerous, more or
less unsuccessful, attacks on RSA. The strongness of RSA bases on
the difficulty to factorize integers as well as to compute the 
discrete logarithm.
For more details, see e.g.\
\cite{Bauer-2000,Buchmann-2001,Cormen-Leiserson-Rivest-1990,Koblitz-1994};
cf.~also
	\href{http://www.math-it.org}{\texttt{\footnotesize http://www.math-it.org}}

\section{RSA cryptosystem}
The RSA cryptosystem, named after its inventors Ron Rivest, Adi Shamir, 
and Len Adleman (1978), was the first public key cryptosystem and is still the 
most important one. It is based on the dramatic difference between the 
ease of finding large prime numbers and computing modular
powers on the one hand, and
the difficulty of 
\emph{factorizing} a product of large prime numbers 
as well as \emph{inverting} 
the modular exponentiation.

Generally, in a public key system, each participant has both a 
\emph{public key} and a \emph{private key}, which is held secret. Each key 
is a piece of information. In the RSA cryptosystem, each key consists of a 
group of integers. The participants are traditionally called 
Alice and Bob, and we denote their public and secret keys as $P_{A}$, 
$S_{A}$ for Alice and $P_{B}$, $S_{B}$ for Bob. All participants 
create their own pair of public and private keys. Each keeps his private 
key secret, but can reveal his public key to anyone or can even publish 
it. It is very convenient that everyone's public key is available in a 
public directory, so that any participant can easily obtain the public 
key of any other participant, just like we nowadays can get anyones 
phone number from a public phone book.

In the \emph{RSA cryptosystem}, each participant creates his public and 
private keys with the following procedure.
\begin{enumerate}
	\item  Select at random two large prime numbers $p$ and $q$, 
	$p$ $\not=$ $q$. 
	(The primes might be more than 200 digits each, i.e.\ more than 
	660 bits.)

	\item  Compute $n=pq$ and the Carmichael function $\lambda(n) = 
	\mathrm{lcm}\,(p-1,q-1).$

	\item  Select an integer $d$ relatively prime to 
	$\lambda(n)$. ($d$ should be of the magnitude of $n$,
	i.e., $d$ $\lessapprox$ $\lambda(n)$.)

	\item  Compute $e$ as the multiplicative inverse of $d$ modulo 
	$\lambda(n)$, such that $ed = 1$ 
	$\mathrm{mod}$ $\lambda(n)$.
	This is done efficiently by the extended Euclidean algorithm.

	\item  Publish the pair $P=(e,n)$ as the \emph{public key}.

	\item  Keep secret the pair $S=(d,n)$ as the \emph{private} 
	or \emph{secret key}.
\end{enumerate}
For this procedure, the domain of the messages is $\mathbb{Z}_{n}$. 
For each participant of a cryptosystem, the four-tuple $(e,d,p,q) \in 
\mathbb{N}^4$ is called \emph{(individual) RSA key system}.
The key parameter $e$ is also called the \emph{encryption exponent}, 
$d$ the \emph{decryption exponent}, and $n$ the \emph{RSA modulus}.

\begin{figure}[ht]
	\centering
	\includegraphics[height=30ex]{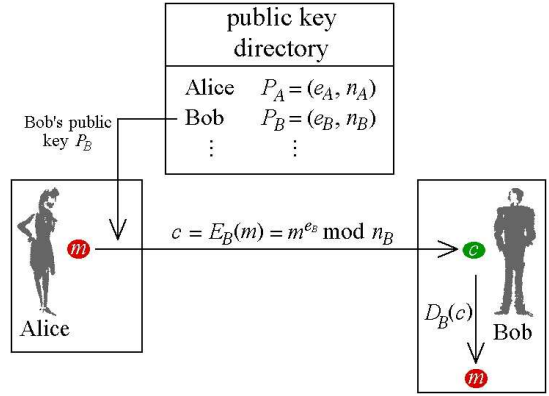}
	\caption{\small 
		Alice sends an encrypted message $m$
		to Bob, using \emph{his} public RSA key $P_B$.
		}
		\label{fig-rsa-encrypt}
\end{figure}

\begin{figure}[ht]
	\centering
	\includegraphics[height=30ex]{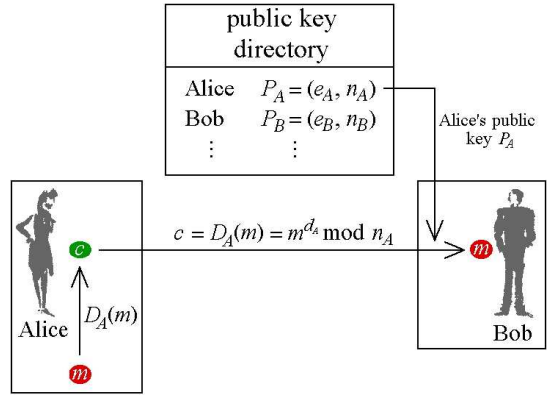}
	\caption{\small 
		Alice sends a digitally signed 
		message to Bob; Bob uses Alice's public key
		to decrypt the message and to verify this way
		that Alice has signed it with her private key.%
		}
		\label{fig-rsa-sign}
\end{figure}

The 
encryption of a message $m\in \mathbb{Z}_{n}$ associated with a public 
key $P=(e,n)$ is performed by the function 
$E: \mathbb{Z}_{n} \to \mathbb{Z}_{n}$,
\begin{equation}
	E(m) = m^e \ \mathrm{mod} \ n.
	\label{E(m)}
\end{equation}
The decryption of a ciphertext $c\in\mathbb{Z}_{n}$ associated with the 
private key $S=(d,n)$ is done by  the mapping 
$D: \mathbb{Z}_{n} \to \mathbb{Z}_{n}$,
\begin{equation}
	D(c) = c^d \ \mathrm{mod} \ n.
	\label{D(m)}
\end{equation}
The procedure where Alice sends an encrypted message to Bob is
schematically shown in Figure \ref{fig-rsa-encrypt}.
A qualitatively new possibility offered by public key systems (and being 
unimplementable with symmetric key systems) is the procedure of
\emph{digital signature}. How an RSA cryptosystem
enables Alice to digitally sign a
message and how Bob can verify that it \emph{is} signed by Alice
is sketched in Figure \ref{fig-rsa-sign}. 
As a matter of course, this verification in fact is possible only
if the authenticity of Alice's public key $P_A$ is guaranteed
such that Bob can assume that it is her key (and not a third
person's one) which he uses. This guarantee is the job of 
so-called trust centers.

The correctness of RSA, i.e., the fact that $E$ and $D$ define inverse 
functions on $\mathbb{Z}_{n}$ ($D\circ E$ $=$ $E \circ D$ $=$ 
$\mathrm{id}_{\mathbb{Z}_{n}}$) relies on the simple fact that
\begin{equation}
	m^{ed} = m \ \mathrm{mod} \ n  
	\qquad \mbox{for } m \in \mathbb{Z}_{n},
	\label{D(E(m))}
\end{equation}
which is immediately proved by the corollary of Carmichael \ref{cor-Carmichael},
p.~\pageref{cor-Carmichael}.
For details see, e.g., 
\cite{Bauer-2000,Buchmann-2001,Cormen-Leiserson-Rivest-1990}.

\begin{bem}
	\label{bem-RSA-Euler}
	Often one finds the definition of RSA cryptosystems based on the 
	Euler function $\varphi$
	rather than on the Carmichael function $\lambda$, cf.\ 
	\cite{Cormen-Leiserson-Rivest-1990}.
	However, since $\varphi(pq)$ $=$ $(p-1)(q-1)$,
	both function values $\varphi(pq)$ and $\lambda(pq)$ 
	share the same divisors. Therefore, a possible key 
	parameter $d$ relatively
	prime to $\lambda(pq)$ is also relatively prime to $\varphi(pq)$, 
	and vice versa.
	Only the resulting counter key $e$ may differ. To be more precise, 
	any possible RSA
	key pair of a system based on the Euler function is a possible key 
	pair with respect
	to the Carmichael function, whereas the reverse is not generally 
	true. 
	(Proof: Since $\lambda(n) | \varphi(n)$, the equality $ed = 1 
	\mbox{ mod } \varphi(n)$
	implies $ed = 1 \mbox{ mod } \lambda(n)$.)
	Using the Euler function $\varphi$, the correctness of RSA
	is shown with the Euler theorem \ref{satz-Euler} on 
	p.~\pageref{satz-Euler}, instead of
	the corollary of Carmichael.
\end{bem}

\subsection{Properties of an RSA key system}
\begin{satz}
	Let $p$, $q \in \mathbb{N}$ be two primes, $p,q>1$, $p\ne q$. 
	Then the number $\nu_{pq}$ 
	of all possible key pairs $(P,S)$ $=$ $((e,pq),(d,pq))$ is given by
	\begin{equation}
		\nu_{pq} = \varphi(\lambda(pq)).
		\label{nu-pq}
	\end{equation}
	The (trivial) keys with $e=d=1$ and with $e=d=\lambda(pq)-1$
	are always possible, and
	\begin{equation}
		2 < \nu_{pq} < {(p-1)(q-1) \over \gcd\,(p-1,q-1)}.
		\label{nu-pq-estimate}
	\end{equation}
\end{satz}
\emph{Proof.} Since $ed = 1$ mod $\lambda(pq)$, 
without restriction to 
generality we have $0<e,d<\lambda(pq)$. Moreover,
gcd$\,(d,\lambda(pq))$ $=$ gcd$\,(e,\lambda(pq))$ $=$ 1, because for 
an arbitrary integer $a$ with gcd$\,(a$, $\lambda(pq))$ $>$ 1 there exists 
no $b\in\mathbb{N}$ such that $ab = 1$ mod $\lambda(pq)$. Therefore,
$e,d \in \mathbb{Z}^{*}_{\lambda(pq)}$.
In turn, to any $a \in \mathbb{Z}^{*}_{\lambda(pq)}$ there exists an
integer $b$ 
such that $ab = 1$ mod $\lambda(pq)$, since $\mathbb{Z}^{*}_{\lambda(pq)}$ 
is a group. But the order of $\mathbb{Z}^{*}_{\lambda(pq)}$ is exactly 
$\varphi(\lambda(pq))$.

It is clear that $1\cdot 1$ $=$ 1 mod $\lambda(n)$, so $e=d=1$ are 
always possible as key parameters.
If $e=d=\lambda(pq)-1$, we have $ed$ $=$ $\lambda^2(pq)$ $-$
$2\lambda(pq)$ $+$ 1 $=$ 1 mod $\lambda(pq)$, so $e$ and $d$ are always 
possible, too. By (\ref{2-lambda-phi}), 
$\lambda(pq)$ is even and (by $pq \geqq 6$)
greater than 2, so $\nu_{pq}>2$. The maximum number of elements on the other 
hand is $\lambda(pq)-1$.
\hfill{$\square$}

\phantom{pause}

The plot of all possible RSA key parameters $(e,d)$  
reveals general symmetries in the $(e,d)$-plane. 
First we observe that if $P=(e,n)$, $S=(d,n)$ is a possible 
RSA key pair, then trivially also $P'=(d,n)$, $S'=(e,n)$ is 
possible, because $ed=de=1$ mod 
$\lambda(n)$. Furthermore,
\begin{figure}[ht]
	\centering
	\includegraphics[height=60mm]{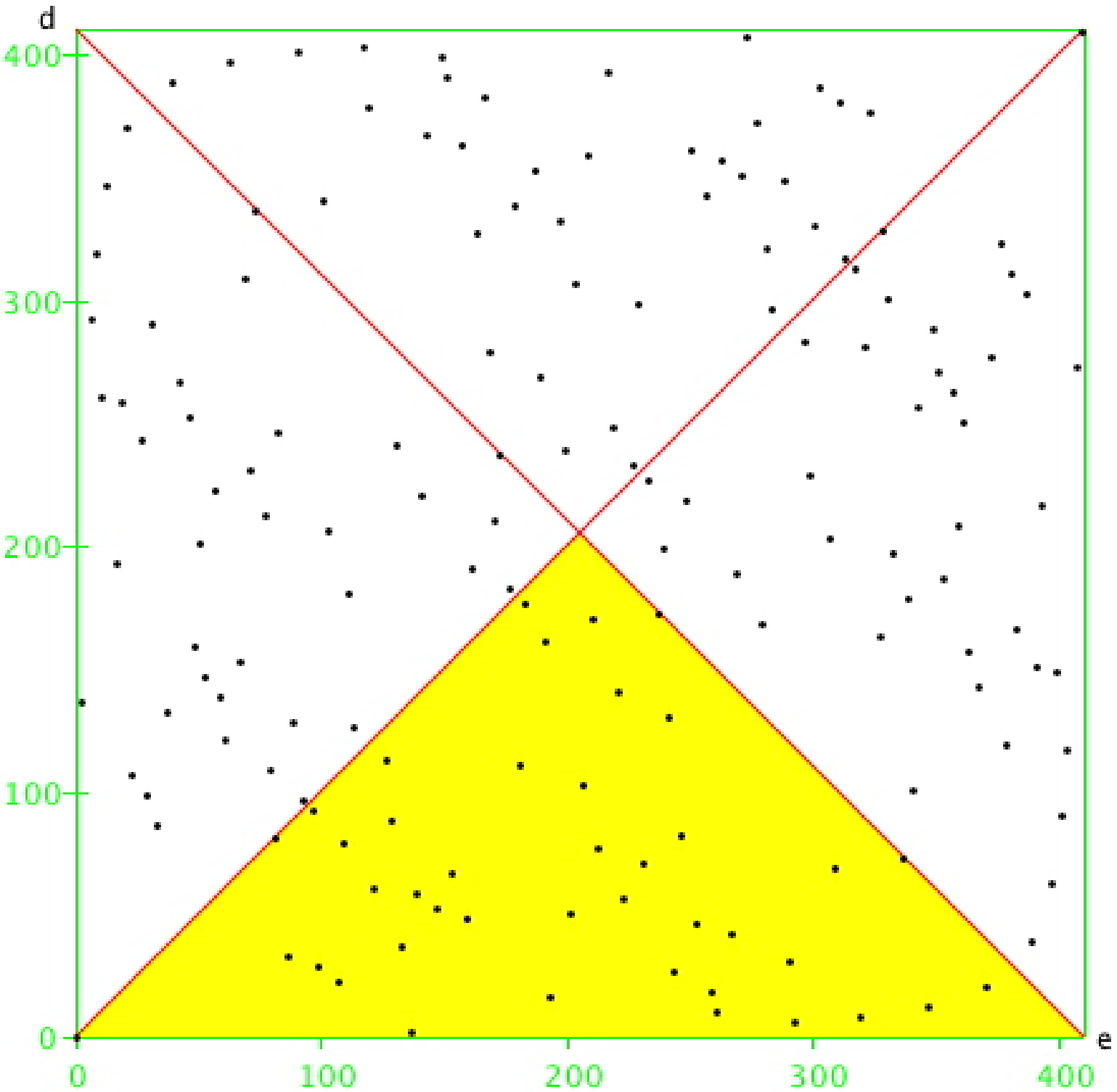}
	\includegraphics[height=60mm]{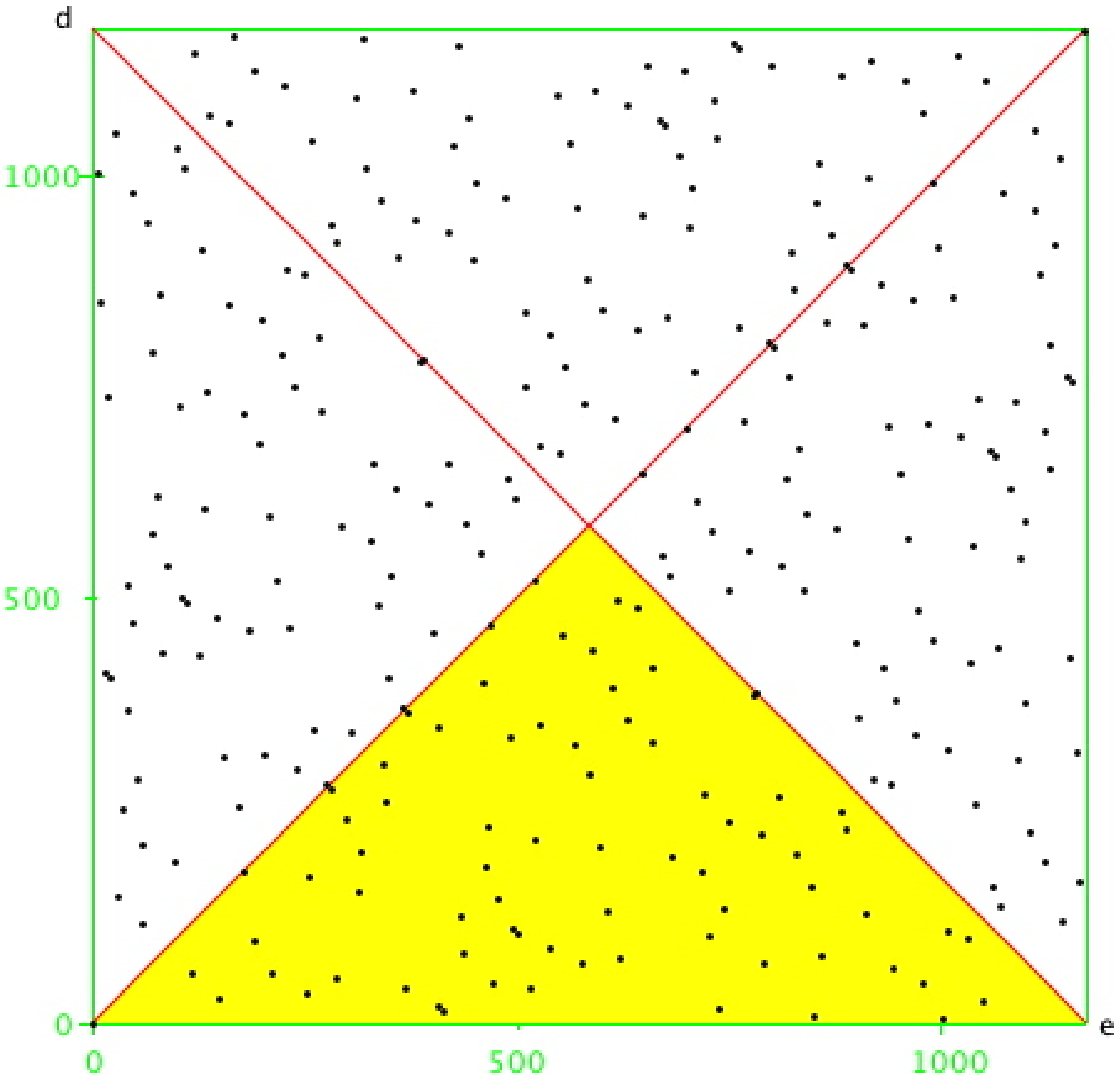}
	\caption{\small 
		Plots of the possible RSA key parameter pairs 
		$(e,d)$ $\in$ $[0,\lambda(n)-1]^2$
		for different primes $p$ and $q$, represented as points 
		in the $(e,d)$-plane. For the first plot, $p=11$ and $q=83$, for the 
		second one $p=19$ and $q=131$. The shaded region is $U$
		as given by (\ref{urcell}).}
		\label{fig-e-d}
\end{figure}
if $ed$ $=$ 1 mod $\lambda(n)$ and $0<e,d<\lambda(n)/2$, then
\begin{equation}
	e' = \lambda(n) - e, \qquad d' = \lambda(n) - d
	\label{e'-d'}
\end{equation}
satisfy $\lambda(n)/2<e',d'<\lambda(n)$ as well as 
$$e'd' = \lambda^2(n) - \lambda(n)(e+d) + ed 
       = ed \ \mathrm{mod} \ \lambda(n).$$
Therefore, $P'=(e',n)$ and $S'=(d',n)$ are possible RSA keys, too.

To sum up, all possible RSA key parameters $(e,d)$, plotted in the 
square lattice $[0,\lambda(n)-1]^2 \subset \mathbb{N}^2$
with edges ranging from 0 to $\lambda(n)-1$, form a pattern which is 
symmetric to both the principal and the secondary square diagonals, 
see Figure \ref{fig-e-d}. Thus, the region
\begin{equation}
	U = \{ (e,d) \in [0, \lambda(n)-1]^2: 
	    0 < d \leqq \min( e, \lambda(n) - e) \}
	\label{urcell}
\end{equation}
contains all information to generate the rest of the square lattice by 
reflections at the main diagonal ($d \leftrightarrow e$) and at the secondary 
diagonal (\ref{e'-d'}).

\subsection{Classical RSA attacks}
There are several specific methods to break an RSA cryptosystem. The 
initial situation for an attack is that an eavesdropper knows the 
public key $P=(e,n)$ and the encrypted message 
$c$
. For details see, e.g., \cite{Bauer-2000} and \cite[\S7]{Buchmann-2001}.

\subsubsection{Factorization of the RSA modulus \textit{n}}
If the eavesdropper succeeds in finding the factorization $n=pq$ of $n$, 
knowing $e$ he can easily compute $d$. But factorization of numbers 
$n=pq$ with
\begin{equation}
	\fbox{$
	p,q > 10^{200}
	$}
	\label{attack-1}
\end{equation}
(hence $n>10^{400}$, i.e., $n$ has length more than about 1320 bits), is 
difficult with current technology, if $p$ and $q$ differ enough,
\begin{equation}
	\fbox{$
	|p-q| > 10^{100}.
	$} 
	\label{attack-2}
\end{equation}
Otherwise $n$ can 
be factorized efficiently by exhaustive search of two integers $n_{+}$ 
and $n_{-}$ satisfying $n$ $=$ $n_{+}^2$ $-$ $n_{-}^2$, beginning at 
$n_{+} = \lceil\sqrt{n}\,\rceil$ and $n_{-}=0$. These two integers then 
necessarily obey 
$ n_{\pm} = {p\pm q \over 2}.$

It can be proved that, knowing the public key $(e,n)$, factorizing the 
RSA modulus $n$ is as difficult as finding the secret key $(d,n)$, see
\cite[\S7.2.5]{Buchmann-2001}.

Factorization is the most efficient known attack on RSA. The fastest
known factorization method, the 
\emph{number field sieve}\index{number field sieve} of John Pollard in 
1988, yields running times for a 10 GHz computer as given
in Table \ref{tab-cpu}.
\begin{table}[htp]
\centering
\begin{tabular}{llll}
	\hline
	magnitude of the number & bits & operations & CPU time \\
	\hline
	$n \approx 10^{50}$ & 167 & $1.4 \cdot 10^{10}$ & 14 seconds \\
	$n \approx 10^{75}$ & 250 & $9 \cdot 10^{12}$   & 2.5 hours \\
	$n \approx 10^{100}$ & 330 & $2.3 \cdot 10^{15}$ & 26.6 days \\
	$n \approx 10^{200}$ & 665 & $1.2 \cdot 10^{23}$ & 3.8 mio years \\
	$n \approx 10^{300}$ & 1000 & $1.5 \cdot 10^{29}$ & 
	  $4.9 \cdot 10^{12}$ years \\
	\hline
\end{tabular}
\caption{\small 
	\label{tab-cpu}CPU times for 
	factorizing of numbers $n$ on a 10 GHz computer.}
\end{table}

\subsubsection{Chosen-plaintext attack}
The eavesdropper systematically encrypts all messages $m$ with Bob's
public key $P_B$ until he achieves the ciphertext $c$. This attack is
efficient if the set of messages $m$ is small or if the message
$m$ is short.
\begin{equation}
	\fbox{\parbox{100mm}{
		``Pad'' each message such that its size is of the magnitude
		of the modulus. 
		Use ``probabilistic encryption,'' where a given plaintext
		is mapped onto several ciphertexts.
	}}
	\label{attack-plain}
\end{equation}

\subsubsection{Chosen-ciphertext attack}
There is a similar method, the chosen-ciphertext attack, which can be applied
if Bob \emph{signs} a document with his private key. The eavesdropper
receiving the ciphertext $c$ and wishing to find the decryption $m=c^d$ mod $n$
chooses a random integer $s$ and asks Bob to digitally sign the innocent-looking
message $\tilde c = s^e c$ mod $n$. From his answer $\tilde m = \tilde c^d$ it is easy
to recover the original message, because $m = \tilde m/s$ mod $n$.
\begin{equation}
	\fbox{\parbox{100mm}{
		Never sign unknown documents; before signing a document, always
		apply a one-way hash function to it.
	}}
	\label{attack-cipher}
\end{equation}

\subsubsection{Message iteration}
Let be $c_{i}\in \mathbb{Z}_{n}$ be iteratively defined as
$$ c_{0}=m, \qquad c_{i} = c_{i-1}^e \ \mathrm{mod}\ n \qquad 
   (i = 1, 2, \ldots).$$
In fact, $c_{i} = m^{e^i}$ mod $n$, and $c_{1}=c$ is the ciphertext.
The smallest index $k$ with 
$c_{k+1} = c_{1}$ is the iteration exponent or period of $m$, cf.\ 
definition \ref{def-iteration}: 
it exactly shows (!) the original message,
$$ c_{k} = m.$$
Such a period $k$ uniquely exists, it is the order of $e$ modulo 
$\lambda(n)$, $k=\mathrm{ord}_{\lambda(n)}(e)$, cf.\ (\ref{order-def}). 
Thus it divides $\lambda(\lambda(n))$ and $\varphi(\lambda(n))$.
To avoid an efficient attack by iteration, $\lambda(\lambda(n))$ and the 
order of $e$ with respect to $\lambda(n)$ have to 
be large,
\begin{equation}
	\fbox{$
	\lambda(\lambda(n)),\ \mathrm{ord}_{\lambda(n)}(e) > 10^{200}. 
	$} 
	\label{attack-3}
\end{equation}
This condition is satisfied for so-called ``doubly safe primes'' 
$p$ and $q$: A prime $p$ is \emph{doubly safe}, if both $(p-1)/2$ and 
$(p-3)/4$ are primes. For instance, 11, 23, 47, 167, 359 are doubly safe 
primes. A doubly safe 
prime $p$ $\not=$ 11 always has the form $24a-1$, or $p$ $=$ $-1$ mod 24. 
For two doubly safe primes $p$, $q$, we have $\lambda(pq)$ 
$=$ 2 ${p-1 \over 2}$ ${q-1 \over 2}$, 
and therefore $\lambda(\lambda(pq))$ $=$ 
$\mathrm{lcm}(2,{p-3 \over 2}$, ${q-3 \over 2})$ $=$
$\frac12$ ${p-3 \over 2}$ ${q-3 \over 2}$ $=$ 
$(p-3)(q-3)/8$.

\subsubsection{Broadcast decryption by the low-exponent attack}
In general, it may be convenient to use a small public key 
parameter $e$ such that the encryption of a message is easy to compute 
(for instance for a small chip card). However, suppose Alice 
sends the same message to $l$ different participants whose public keys 
are $P_{i}=(e,n_{i})$ where the $n_{i}$'s are relatively prime to each 
other and $l\leqq e$; to emphasize, the public keys have the same 
encryption exponent 
$e$. If an eavesdropper 
receives the $l$ ciphertexts $c_{i}'$ $=$ $m^{e}$ mod $n_{i}$, he can 
easily compute $c'=c_{i}'$ mod $n_{1}\cdots n_{l}$ by the Chinese 
remainder theorem. But if the product $n_{1} \cdots n_{l}$ is great 
enough, this is the same as $c' = m^e$. This equation is invertible, viz.,
$ m = \sqrt[e]{c'},$
and the original message is computed. 
To avoid this attack, each pair of public keys $P_{i}=(e_{i},n_{i})$
$P_{j}=(e_{j},n_{j})$ and any broadcast message $m$ must satisfy
\begin{equation}
	\fbox{$
	e_{i} \not= e_{j} \quad \mbox{ or } \quad
	m^{e_{i}},\ m^{e_{j}} > n_{i}n_{j} 
	$} 
	\label{attack-4}
\end{equation}

\subsubsection{Broadcast decryption by the common modulus attack}
If a plain text $m$ is encrypted twice by the RSA system 
using two public keys $P_{i}$ $=$ $(e_{i},n)$, $i=1,2$, with a common 
modulus $n$ and gcd$\,(e_{1},e_{2})$ $=$ 1, then $m$ can be recovered
efficiently from the two ciphertexts $c_{1}$ and $c_{2}$,
each of which given by 
$c_{i}$ $=$ $m^{e_{i}}$ mod $n$.
This is done by the following procedure. 
\begin{enumerate}
	\item Compute $x_{1}$, $x_{2}$ satisfying 
	$x_{1}e_{1}$ $+$ $x_{2}e_{2}$ $=$ 1 by the extended 
	Euclidean algorithm, where the indices are chosen such that
	$x_2<0$.
	
	\item Determine $y$ 
	satisfying 1 $=$ $yc_{2}$ $+$ $kn$ by the extended Euclidean 
	algorithm.

	\item Calculate 
	${c_{1}}^{x_{1}}{y}^{-x_{2}}$ --- this is the plain text! 
\end{enumerate}
The reason is that
${c_1}^{x_1} y^{-x_2}$ $=$
${c_{1}}^{x_{1}}$ ${c_{2}}^{x_{2}}$ $=$ $m^{x_{1}e_{1} + x_{2}e_{2}}$ $=$ $m$ 
mod $n$.
E.g., let be $P_{1}$ $=$ $(3,493)$ and $P_{2}$ $=$ $(5,493)$, 
and the corresponding ciphertexts $c_{1}$ $=$ 293 and $c_{2}$ $=$ 421. 
Then the extended Euclidean algorithm yields $x_1$ $=$ $2$ and 
$x_2$ $=$ $-1$, and thus $y=89$ and $k=-76$ (such that 89 $\cdot$ 421
$-$ 76 $\cdot$ 493 $=$ 1); 
finally,  $293^{2}$ $\cdot$ $89^{1}$ $=$ 67 $\cdot$ 89 $=$ 5963 $=$ 47 
mod $n$, i.e.\ $m$ $=$ 47 is the plaintext. In fact, 493 $=$ 17 $\cdot$ 29,
and $S_{1}$ $=$ (17, 29, 75), $S_{2}$ $=$ (17, 29, 45), and 
$m$ $=$ $c_{1}^{75}$ $=$ $c_{2}^{45}$ $=$ 47 mod 493. 

Therefore, to avoid common modulus attacks, a sender should regard:
\begin{equation}
	\fbox{\parbox{100mm}{
		Never send identical messages to receivers 
		with the same modulus and relatively prime 
		encryption exponents.
	}}
	\label{attack-5}
\end{equation}

\section{The Euler function ray attack}

\subsection{The $\omega$-function and the order of a number modulo \textit{n}}
\begin{defi}
	Let be $n\in \mathbb{N}$, $n>1$, and $\mathbb{Z}_{n}^{*}$ the 
	multiplicative group 
	modulo $n$. Then the \emph{order} ord$_{n}(m)$ of $m \in \mathbb{Z}_{n}^{*}$ 
	is given by 
	\begin{equation}
		\mathrm{ord}_{n}(m) = \min\, \{k \in \mathbb{N}:\ k>0,\ m^k = 1 \ 
		  \mathrm{mod}\ n\}.
		\label{order-def}
	\end{equation}
	If gcd\,$(m,n)>1$, ord$_{n}(m) = \infty$.
\end{defi}
Let $\langle m \rangle$ denote the subgroup of $\mathbb{Z}_{n}^{*}$ generated by 
$m$. E.g., $\langle 2 \rangle$ = \{1, 2, 4\} in $\mathbb{Z}_{7}$, and 
ord$_{7}(2)$ = 3. Note that $\varphi(7)$ = $\lambda(7)$ = 6.

\begin{lem}
	Let be $m,n \in \mathbb{N}$, with 
	$\mathrm{gcd}(m,n) = 1$ and $m<n$. Then
	\begin{equation}
		\mathrm{ord}_{n}(m)\mid\lambda(n).
		\label{order-lambda}
	\end{equation}
	Moreover,
	\begin{equation}
		\lceil \log_{m} n \rceil \leqq \mathrm{ord}_{n}(m) 
		\leqq \lambda(n) \leqq n-1.
		\label{log-ord-lambda}
	\end{equation}
\end{lem}
\emph{Proof.} With Carmichael's 
theorem \ref{satz-Carmichael} and with the Lagrange theorem 
\cite[\S33]{Cormen-Leiserson-Rivest-1990} equation 
(\ref{order-lambda}) is deduced.

Let $a=\mathrm{ord}_{m}(n)$. Since $m>1$, we have $m^a>n$ 
to obtain $m=1 \mbox{ mod }n$. This implies $a>\log_{m} n$. 
The upper limits follow from the relations (\ref{lambda-estimate}) 
and (\ref{order-lambda}).
\hfill{$\square$}

\begin{defi}
	\label{def-iteration}
	Let be $m,n,e \in \mathbb{N}$, $n>1$, and define the sequence 
	($c_{0}$, $c_{1}$, $c_{2}$, \ldots) 
	iteratively by
	\begin{equation}
		c_{0}=m, \qquad c_{i} = c_{i-1}^e \ \mathrm{mod}\ n 
		\qquad (i=1,2,\ldots).
		\label{c-i}
	\end{equation}
	Then the smallest $k$ $\geqq$ 1 such that $c_{k}$ $=$ $c_{0}$ is 
	called 
	$(n,e)$-\emph{iteration exponent} $s(n$, $e$, $m)$ of $m$. It is 
	the period 
	of the \emph{cycle} ($c_{0}$, $c_{1}$, \ldots, $c_{s(n,e,m)-1}$) 
	to which $m$ 
	belongs. A cycle with period one is a fixed point.
\end{defi}
\begin{lem}
	Let be $e,m,n$ and the sequence $(c_{0},c_{1},c_{2},\ldots)$ 
	as in definition \ref{def-iteration}. Let moreover be $e$ relatively 
	prime to $\lambda(n)$.
	Then the $(n,e)$-iteration exponent $s(n,e,m)$ satisfies
	\begin{equation}
		s(n,e,m) \mid \lambda(\lambda(n)). 
		\label{iteration-lambda}
	\end{equation}
\end{lem}
\emph{Proof.} Note that for the sequence (\ref{c-i}) we have $c_{i} = 
m^{e^i}$ mod $n$. For $s(n,e,m)$ we thus have
\begin{equation}
	m^{e^{s(n,e,m)}} = m^e \ \mathrm{mod}\ n.
	\label{m^i^s}
\end{equation}
By (\ref{equ-Carmichael}) we have $e^{s(n,e,m)} = e$ mod $\lambda(n)$,
which implies by definition \ref{def-iteration}
that ord$_{\lambda(n)}(e)$ = 
$s(n,e,m).$ Relation (\ref{order-lambda}) yields the assertion.
\hfill{$\square$}

\begin{bsp}
	Let be $e=7$, $n=55=5\cdot 11$. Then we have $\lambda(55)=20$, and
	$\lambda(\lambda(55))=4.$ Denoting $c_{0}=51$, we obtain
	\begin{eqnarray*}
		c_{1} & = & 51^7 \ \mathrm{mod}\ 55 =  6 \\
		c_{2} & = &  6^7 \ \mathrm{mod}\ 55 = 41 \\
		c_{3} & = & 41^7 \ \mathrm{mod}\ 55 = 46 \\
		c_{4} & = & 46^7 \ \mathrm{mod}\ 55 = 51 = c_{0}
	\end{eqnarray*}
	Hence, the period of the cycle which 51 belongs to is $s(n,e,m) = 4.$ 
	Note by (\ref{iteration-lambda}) that this is the maximum value.
	Analogously, there are the following cycles.
	\begin{center}
		\begin{tabular}{lll}
			\hline
			9 fixed points &%
			  \multicolumn{2}{l}
			  {(0), (1), (10), (11), (21), (34), (44), (45), (54)} 
			  \\[1ex]
			3 cycles of period 2 &%
			  \multicolumn{2}{l}
			  {(12, 23), \ (22, 33), \ (32,43)} \\[1ex]
			10 cycles of period 4 
			    &( 2, 18, 17, \ 8), &( 3, 42, 48, 27)  \\
			    &( 4, 49, 14, \ 9), &( 5, 25, 20, 15)  \\
			    &( 6, 41, 46,  51), &( 7, 28, 32, 13)  \\
			    &(16, 36, 31,  26), &(19, 24, 29, 39)  \\
			    &(30, 35, 40,  50), &(37, 38, 47, 53)  \\
			\hline
		\end{tabular}
	\end{center}
\end{bsp}

\begin{defi}
	Let be $m,n \in \mathbb{Z}$, $n\geqq 0$. Then we define the function
	\begin{equation}
		\omega_{m}(n) = \left\{ \begin{array}{ll}
		  \mathrm{ord}_{n}(m) & \mbox{if } \mathrm{gcd}\,(m,n) = 1,\\[1.0ex]
		  0                   & \mbox{if } \mathrm{gcd}\,(m,n) \not= 1.
		\end{array} \right.
		\label{def-omega}
	\end{equation}
\end{defi}
It is obvious that $m^{\omega_{m}(n)}$ $=$ 1 mod $n$ for any 
$m,n\in\mathbb{Z}$, $n\geqq 0$ (since this is the definition of the 
order function). Substituting $n$ by $\omega_{m}(n)$ immediately yields
\begin{equation}
	m^{\omega_{m}(\omega_{m}(n))} = 1 \ \mathrm{mod}\ \omega_{m}(n).
	\label{omega(omega)}
\end{equation}
Here ``$a$ $=$ $b$ mod 0'' has to be understood as a congruence in 
$\mathbb{Z}$, i.e.\ as ``$a$ $=$ $b$.''
By iteration, we obtain the cascading-$\omega$ equation
\begin{equation}
	m^{\omega_{m}^{(r)}(n)} = 1 \ \mathrm{mod}\ \omega_{m}^{(r-1)}(n),
	\qquad \mbox{where } r\geqq 1.
	\label{omega-cascading}
\end{equation}
where $\omega_{m}^{(r)}(n) = 
\omega_{m}(\omega_{m}(\ldots(\omega_{m}(n))\ldots))$ denotes the 
$r$-fold composition of $\omega_{m}$.

\begin{satz}
	\label{theo-attack-d=e}
	Let be $d,e,n \in \mathbb{N}$, such that $n>1$, 
	$\mathrm{gcd}\,(e,n)=1$, and $d\cdot e$ $=$ $1$ 
	$\mathrm{mod}$ $\lambda(n)$.
	Then $\omega_{e}(\omega_{e}(n)) > 0$, and
	\begin{equation}
		d = e^{\omega_{e}(\omega_{e}(n))-1} \ \mathrm{mod}\ \omega_{e}(n).
		\label{attack-d=e}
	\end{equation}
\end{satz}
\emph{Proof.} First we note by (\ref{order-lambda}) that
$\omega_{e}(n)$ $|$ $\lambda(n)$. Therefore, $de = 1$ mod $\lambda(n)$ implies
\begin{equation}
	d\cdot e = 1 \ \mathrm{mod} \ \omega_{e}(n).
	\label{de=1}
\end{equation}
(If $de-1$ = $k\lambda(n)$ for a $k\in\mathbb{Z}$, then $de-1$ = 
$k'\omega_{e}(n)$, where $k'$ $=$ $k\lambda(n)/\omega_{e}(n)$.) 
If we had now $\omega_{e}(\omega_{e}(n))$ $=$ 0, then $e$ would divide 
$\omega_{e}(n)$ and hence $\lambda(n)$: But then there 
would be no $d$ with $de$ $=$ 1 $\mathrm{mod}$ $\lambda(n)$. Hence, 
$\omega_{e}(\omega_{e}(n))>0$.
Moreover, by the cascading-$\omega$ equation (\ref{omega(omega)}) we have
\begin{equation}
	e^{\omega_{e}(\omega_{e}(n))-1}\cdot e = 1 \ \mathrm{mod} \ \omega_{e}(n).
	\label{ee^o=1}
\end{equation}
Equation (\ref{attack-d=e}) follows immediately from (\ref{de=1}) and (\ref{ee^o=1}).
\hfill{$\square$}

\begin{bsp}
	Let be $n=221$ and $e=11$. Then $\omega_{11}(221)$ $=$ 48, 
	$\omega_{11}(48)$ $=$ 4, hence
	$$d = 11^3 = 35 \ \mathrm{mod}\ 48.$$
	Therefore, the possible $d<221$ are $d$ $=$ 35, 83, 131, 179. In fact, 
	221 $=$ 13 $\cdot$ 17, and $\lambda(221)$ $=$ 48; this means that $11 
	\cdot 35$ $=$ 1 mod $\lambda(221)$, or $d=35$.
\end{bsp}

The two shoulders on which Theorem \ref{theo-attack-d=e} rests are equations 
(\ref{de=1}) and (\ref{ee^o=1}). They can be extended to analogues 
for the following corollary. 

\begin{cor}
	\label{cor-attack-d=e}
	Let be $e,n, a,b \in \mathbb{N}$ such that $n>1$ and 
	$\mathrm{gcd}\,(e,n)=1$, 
	as well as $\lambda(n)$ $|$ $\omega_{e}(a\omega_{e}(n))$. 
	Then the integer 
	\begin{equation}
		\tilde{d} = e^{b\omega_{e}(a\omega_{e}(n))-1} \ 
		\mathrm{mod}\ a\omega_{e}(n)
		\label{attack-d=e'}
	\end{equation}
	satisfies $\tilde de = 1$ $\mathrm{mod}$ $a\omega_{e}(n)$, 
	and for any number 
	$m$ $\in$ $\mathbb{Z}_{n}$ we have
	\begin{equation}
		m^{e\tilde d} = m \ \mathrm{mod}\ n.
		\label{attack-m^ed}
	\end{equation}
	If the integer $a$ is such that 
	$\omega_{e}(a\omega_{e}(n))$ $|$ $\lambda(n)$, 
	then the unique $d<\lambda(n)$ with 
	$de = 1 $ $\mathrm{mod}$ $\lambda(n)$ is related to 
	$\tilde{d}$ by 
	\begin{equation}
		d = \tilde{d} \ \mathrm{mod}\ a\omega_{e}(n).
		\label{attack-d=d'}
	\end{equation}
\end{cor}
\emph{Proof.} 
Substituting $n$ by  $a\omega_{e}$, from $e^{\omega_{m}(n)}$ $=$ 1 mod $n$ for any $m\in\mathbb{Z}$ we deduce that $e^{b\omega_{e}(a\omega_{e}(n))}$ $=$ 1 mod $a\omega_{e}(n)$. Especially, with (\ref{attack-d=e'}) we have
\begin{equation}
	\tilde{d} \cdot e = e^{b\omega_{e}(a\omega_{e}(n))-1}\cdot e 
		= 1 \ \mathrm{mod} \ a\omega_{e}(n).
	\label{ee^o=1'}
\end{equation}
If  $\lambda(n)$ $|$ $\omega_{e}(a\omega_{e}(n))$, we have 
$m^{\tilde d e \ \mathrm{mod} \ a\omega_{e}(n)}$ $\mathrm{mod}$ 
$n$ $=$ 
$m^{1 \ \mathrm{mod} \ a\omega_{e}(n)}$ $\mathrm{mod}$ $n$ $=$ 
$m^{1 \ \mathrm{mod} \ \lambda(n)}$ $\mathrm{mod}$ $n$ $=$ $m$ $\mathrm{mod}$ $n$.
(Note that $\lambda(n)$ enters the scene in the second last equation to fulfill 
the equation for \emph{all} $m$!) 
In turn, if $\omega_{e}(a\omega_{e}(n))$ $|$ $\lambda(n)$, then
$\tilde de$ $=$ 1 $\mathrm{mod}$ $\lambda(n)$ implies
$\tilde de$ $=$ 1 $\mathrm{mod}$ $a\omega_{e}(n)$; thus (\ref{attack-d=d'}) follows 
from (\ref{ee^o=1'}).
\hfill{$\square$}

\begin{bsp}
	Let be $n=143$ and $e=47$. Then $\omega_{47}(143)$ $=$ 20, and with 
	$a=2$, $b=3$,
	we have $3\omega_{47}(40)$ $=$ 12, hence
	$$d = 47^{11} = 23 \ \mathrm{mod}\ 40.$$
	Therefore, $m^{ed}$ $=$ $m^{1081}$ $=$ $m$ mod 143. In fact, 
	143 $=$ 11 $\cdot$ 13, and $\lambda(143)$ $=$ 60; this means that $47 
	\cdot 23$ $=$ 1 mod $\lambda(143)$, or $d=23$.
\end{bsp}

\begin{bem}
	Given two relatively prime integers $e$ and $n$,
	corollary \ref{cor-attack-d=e} enables us to choose an (almost) 
	arbitrary 
	multiple of the order $\mathrm{ord}_{n}(e)>0$ to find an integer $d$ 
	being a kind of 
	``inverse'' of $e$: If the
	multiple is small enough such that it divides $\lambda(n)$, our 
	result 
	supplies a list of values, one of which satisfies $ed=1$ mod 
	$\lambda(n)$; 
	if the multiple is also a multiple of $\lambda(n)$, we can compute 
	$\tilde{d}$ such that $\tilde{d}e$ $=$ 1 mod 
	$a\,\mathrm{ord}_{n}(e)$.	 In particular, by (\ref{2-lambda-phi}) 
	and 
	(\ref{order-lambda}) the Euler function is a multiple of both $\lambda(n)$
	and $\mathrm{ord}_{n}(e)$.
\end{bem}

\subsection{Properties of composed numbers \textit{n = pq}}
Let be $p$, $q$ be two primes, $p\not= q$. Then $n$ $=$ $pq$ is an integer composed  of two primes.
Among the integers $n$ less than 50 there are 13 ones composed of two 
primes, $n=pq$, whereas less than 100 there are 30 ones,
shown in the following tables. 
\begin{center}
	\begin{tabular}{r|*{15}{c}}
		\hline
		$n$ & 
		  6 & 10 & 14 & 15 & 21 & 22 & 26 & 33 & 34 & 35 & 38 & 39 &
		 46 & 51 & 55 \\
		\hline
		$\varphi(n)$ & 
		  2 &  4 &  6 &  8 & 12 & 10 & 12 & 20 & 16 & 24 & 18 & 24 &
		 22 & 32 & 40\\
		$\lambda(n)$ & 
		  2 &  4 &  6 &  4 &  6 & 10 & 12 & 10 & 16 & 12 & 18 & 12 &
		 22 & 16 & 20\\
		\hline
	\end{tabular}
\end{center}
\begin{center}
	\begin{tabular}{r|*{15}{c}}
		\hline
		$n$ & 
		  57 & 58 & 62 & 65 & 69 & 74 & 77 & 82 & 85 & 
		  86 & 87 & 91 & 93 & 94 & 95 \\
		\hline
		$\varphi(n)$ & 
		  36 & 28 & 30 & 48 & 44 & 36 & 60 & 40 & 64 & 
		  42 & 56 & 72 & 60 & 46 & 72 \\
		$\lambda(n)$ & 
		  18 & 28 & 30 & 12 & 22 & 36 & 30 & 40 & 16 &
		  42 & 28 & 12 & 30 & 46 & 36 \\
		\hline
	\end{tabular}
\end{center}
Let us now study the geometric structure of the Euler function.

\begin{satz}
	\label{theo-phi}
	Let $n=pq$ be a positive integer, composed of two primes 
	$p$ and $q$ with $p<q$. 
	For any integer $p_{\mathrm{min}}\in\mathbb{N}$ 
	satisfying
	$p_{\mathrm{min}}$ $\leqq$ $p$ we then have
	\begin{equation}
		\varphi(n) \geqq (p_{\mathrm{min}} - 1) 
		\left( {n \over p_{\mathrm{min}}} - 1 \right).
		\label{phi-estimate}
	\end{equation}
	The inequality is strict, if $p_{\mathrm{min}}<p,q.$
\end{satz}
\emph{Proof.} 
We have 
$\varphi(n)$ $=$ 
$\displaystyle (p-1) \left({n \over p}-1 \right)$, and 
$\varphi(n)$ is a function of $p$:
$$g(p) = \varphi(n) = n - p - {n \over p} + 1.$$
Since $g'(p) = -1+ n/p^2<0$, for fixed $n$ the function $g$ is 
strictly decreasing with respect to $p$, as long as $p<q$, i.e.\ as 
$n/p^2>1$.
\hfill{$\square$}

\phantom{pause}

\begin{figure}[ht]
	\centering
	\includegraphics[height=60mm]{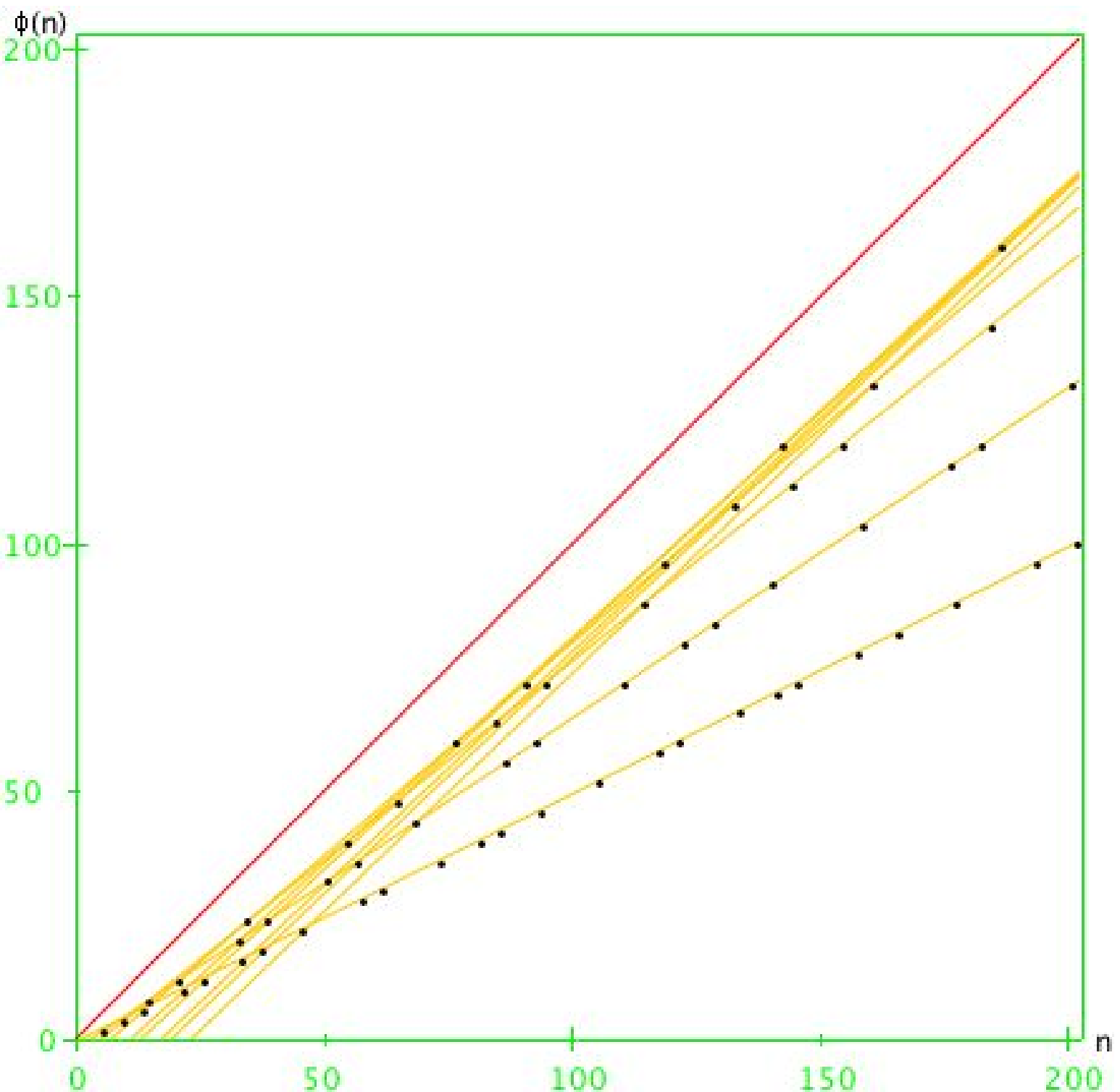}
	\includegraphics[height=60mm]{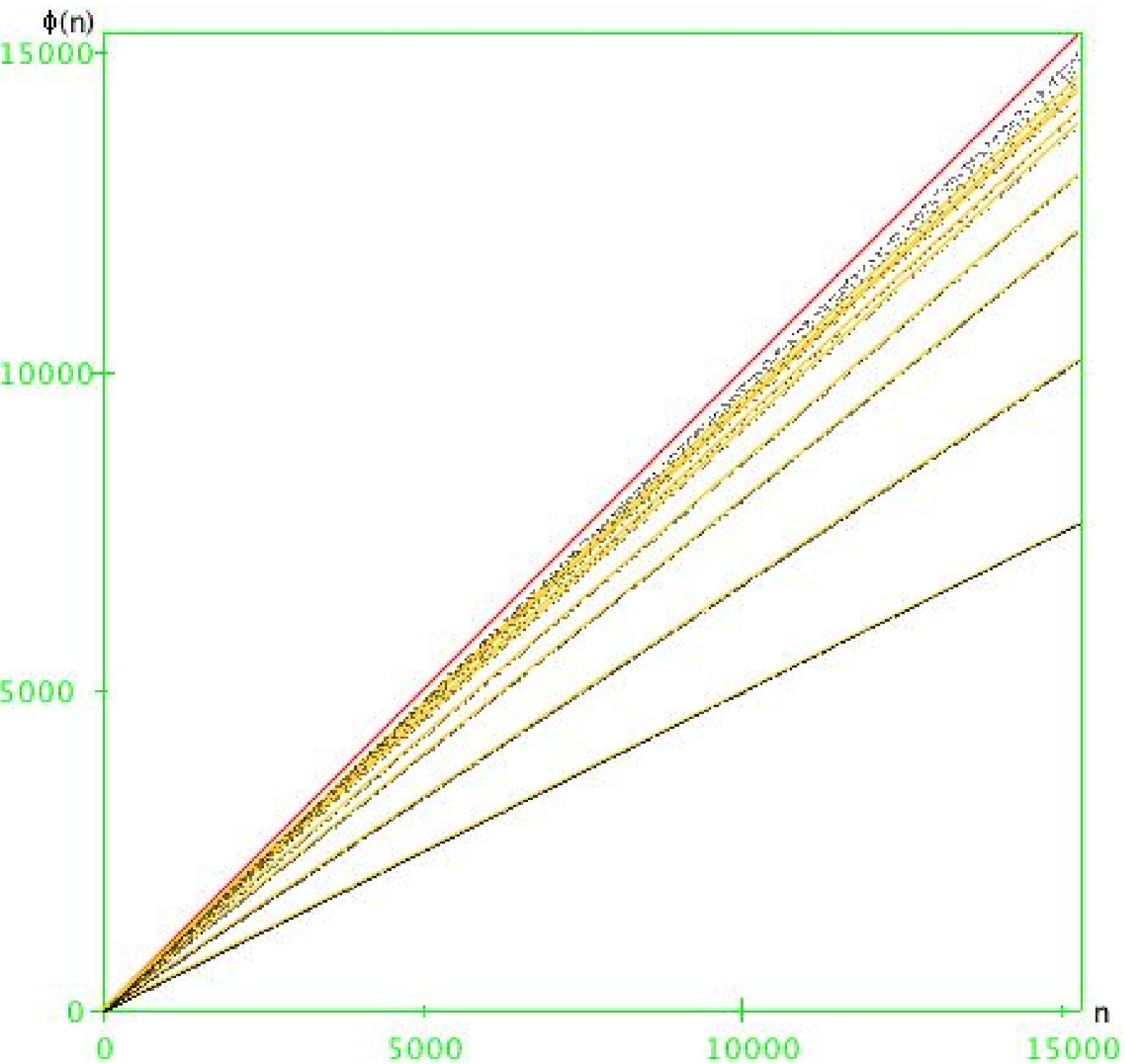}
	\caption{\small 
		Plot of the Euler function 
		$\varphi(pq)$, with $p$, $q$ prime; also sketched are the rays 
		$f_p$ for $p$ $=$ 2, 3, 5, 7, 11, 13, 17, 19, 23.}
		\label{fig-phi-2}
\end{figure}

Geometrically, this result means that in the graph of $\varphi(n)$ the point $(n,\varphi(n))$ lies above the ``Euler function ray''
(see Figure \ref{fig-phi-2})
\begin{equation}
	f_{p}(x) = \left( x, (p-1) \left( {x \over p} - 1 \right) \right).
	\label{f-ray}
\end{equation}

\begin{satz}
	\label{theo-guess}
	Let be $p$, $q$ two primes $p<q$, $e$ an integer with $e>1$,
	and $n$ $=$ $pq$.
	Moreover define for $a$ $\in$ $\mathbb{N}$ the exponents
	$\delta_{e,n}$, $\gamma_{e,a}$ $\in$ $\mathbb{N}$ by
	\begin{equation}
		\delta_{e,n} = \max \{ i \in \mathbb{N}: \ e^i \leqq n \}
		   = \left\lfloor{\ln n \over \ln e}\right\rfloor, \qquad
		\gamma_{e,a} = \max \{ i \in \mathbb{N}: \ e^i \mid a \},
	\label{delta-gamma}
	\end{equation}
	as well as
	\begin{equation}
		r_\pm = \frac12 \left( \Delta \pm \sqrt{\Delta^2 - 4n} \right)
		\qquad \mbox{with }
		\Delta = p+q+\delta_{e,n}.
	\label{r+-}
	\end{equation}
	Then for any integers $b$, $r\in\mathbb{N}$,
	$r_- \leqq r \leqq p$ or $q \leqq r \leqq r_+$, satisfying
	\begin{equation}
		be^{\lfloor f(r) \rfloor} 
		 = 1 \ \mathrm{mod}\ n, \qquad \mbox{where}
		 \quad f(r) = (r-1) \left( \frac{n}{r} - 1 \right),
	\label{guess-r}
	\end{equation}
	the Euler function value $\varphi(n)$ can be computed by
	\begin{equation}
		\varphi(n) = \gamma_{e,b} + \lfloor f(r) \rfloor
	\label{guess-phi}
	\end{equation}
\end{satz}
\emph{Proof.}
Note first that real values for $r_\pm$ always exist since the
term in the square root is positive,
$\Delta^2$ $>$ $(p+q)^2$, i.e.\ $\Delta^2 - 4n$ $>$ $(q-p)^2$ $>$ 0.
We see that
$(p-1) (q-1) - (r-1)\left( {n / r} - 1 \right)$
$=$ $\frac{1}{r}( r^2 - \Delta r + n )$. Solving this quadratic equation
with respect to $r$,
straightforward calculation thus shows that the inequalities for $r$ 
are equivalent to the inequalities
\begin{equation}
	0 \leqq (p-1) (q-1) - (r-1)\left( {n \over r} - 1 \right) \leqq 
	\delta_{e,n},
\end{equation}
which means that $0 \leqq \varphi(n) - (r-1)\left( {n \over r} - 1 \right) 
\leqq \delta_{e,n}.$ 
On the other hand, $b$ being the multiplicative inverse of
$e^r$ by the modular equation in (\ref{guess-r}), we have $b$ $=$ $e^j$ 
mod $n$ for some $j$ $\in$ $\mathbb{N}$, in particular for $j$ $=$ 
$\varphi(n)-r$.
But if $j$ $<$ $\delta_{e,n}$, we have
$b$ $=$ $e^j$, and $j$ $=$ $\gamma_{e,b}$.
\hfill{$\square$}

\begin{bsp}
	\label{ex-guess}
	Let be $p=11$, $q=13$, and $e=7$. Then $\delta_{7,143}$ $=$ 2, and 
	thus $\Delta$ $=$ 26, $r_\pm$ $=$ 13 $\pm$ $\sqrt{26}$. So	$r$ shall satisfy $8\leqq r \leqq 11$ or $13\leqq r \leqq 18$. For 
	$r=8$, e.g., we have
	$$(r-1)\left( {n \over r} - 1 \right) = 7 \cdot 16.875 = 118.125;$$
	Since
	$7^{118} = 108 \ \mathrm{mod}\ 143$, we achieve by the extended 
	Euclidean algorithm
	$b=49=7^2$ (because 1 $=$ $49\cdot 108 - 37 \cdot 143$), and with
	$\gamma_{7,49}=2$ we obtain
	$$ \varphi(143) = 118 + \gamma_{7,49} = 120.$$
	In fact, $\varphi(143) = 10 \cdot 12.$
\hfill{$\square$}
\end{bsp}
\begin{bsp}
	\label{ex-guess-gross}
	Let be $p$ $=$ 3\,336\,670\,033, 
	$q$ $=$ 9\,876\,543\,211, and $e$ $=$ 2. Then 
	$$n = 32\,954\,765\,761\,773\,295\,963,$$
	$\delta_{2,n}$ $=$ 64, and thus 
	$$\Delta= 13\,213\,213\,308, \quad
	r_- = 3\,336\,670\,000.3, \quad
	r_+ = 9\,876\,543\,307.6.$$
	For $r$ $=$ 9\,876\,543\,308, 
	e.g., we have
	$$ i = (r-1)\left( {n \over r} - 1 \right) = 
	  32\,954\,765\,748\,560\,082\,656.$$
	Since
	$$2^{i} = 7\,542\,048\,005\,965\,299\,043 \ \mathrm{mod}\ n,$$ 
	we achieve by the extended Euclidean algorithm
	$$ b=18\,446\,744\,073\,709\,551\,616$$ 
	and with
	$\gamma_{2,b}=64$ we obtain
	$$ \varphi(n) = i + \gamma_{2,b} = 32\,954\,765\,748\,560\,082\,720.$$
\hfill{$\square$}
\end{bsp}

The following lemma tells us the grade of ``coarse graining,'' i.e., a step-width 
that a systematic and definite search for an appropriate Euler function ray 
factor $r$ must use.

\begin{lem}
	\label{lem-coarse-grain-2}
	Let $p$, $q$ be two primes, $p<q$, $e$ an integer $e>1$, and $n=pq$. 	Moreover let 
	$r_+$ and $\delta_{e,n}$ be defined as
	in theorem \ref{theo-guess} by equations (\ref{delta-gamma}) and 	(\ref{r+-}).
	Then 
	\begin{equation}
		r_+ - q > 
		{\delta_{e,n} \over 2}.
	\label{coarse-grain-2}
	\end{equation}
	Moreover,
	\begin{equation}
		p - r_- > \frac{\delta_{e,n}}{2}
		\qquad \mbox{if } \ 
		\textstyle
		\delta_{e,pq} < \frac{2}{3} ( 3p-q).
	\label{coarse-grain-3}
	\end{equation}
\end{lem}
\emph{Proof.}
By
$\Delta^2$ $-$ $4pq$ $=$ $(q-p)^2$ $+$ $2 (p+q) \delta_{e,pq}$ $+$
$\delta_{e,pq}^2$
we achieve for $\delta_{e,pq}>0$ 
\begin{eqnarray*}
	r_+ &=& \frac12 \left( \Delta + \sqrt{\Delta^2-4pq} \right)
		 =  \frac12 \left( \Delta + \sqrt{(q-p)^2 + 2 (p+q) \delta_{e,pq} 
			+ \delta_{e,pq}^2}\ \right) \\
		&>&
		\frac12 \left( \Delta + q-p \right)
		 = \frac12 \left( 2q + \delta_{e,pq} \right)
		 = q + {\delta_{e,pq} \over 2}.
\end{eqnarray*}
Analogously, by 
(\ref{coarse-grain-3}) we have
$2 (q - p) + \frac{3}{2} \delta_{e,pq} < q + p$, i.e.\
$(q-p)^2 + 2(q+p)\delta_{e,pq} + \delta_{e,pq}^2 > 
 (q-p)^2 + 4(q-p)\delta_{e,pq} + 4\delta_{e,pq}^2 
 = (q-p+2\delta_{e,pq})^2$, i.e.
\begin{eqnarray*}
	r_- &=& \frac12 \left( \Delta - \sqrt{(q-p)^2 
	         + 2 (p+q) \delta_{e,pq} + \delta_{e,pq}^2}\ \right)
	\\
	    &<& \frac12 \left( \Delta - \sqrt{(q-p)^2 + 4 (p-q) \delta_{e,pq} 
			+ 4 \delta_{e,pq}^2}\ \right)
	\\
		&=& 
		\frac12 \left( \Delta - q + p - 2 \delta_{e,pq} \right)
		 = p - \frac{\delta_{e,pq}}{2}.
\end{eqnarray*}
\hfill{$\square$}

\subsection{The algorithm}
An algorithm to break an RSA cryptosystem is shown below in pseudocode. It
is invoked with the public key $(e,n)$ and the estimate $r$
for the Euler function ray as input parameters and returns a possible private RSA key parameter $d$ corresponding to $e$. If it fails, 
$d \leqq 0$ is returned.
\begin{footnotesize}
\begin{verbatim}
  long rayAttack ( e, n, r ) {
    // store an array a such that a[i] = m^(2^i) < n:
    a[0] = e;
    j = 1;
    while ( a[j-1] < n ) {
       a[j] = a[j-1] * a[j-1];
       j++;
    }
    delta = 0;
    while ( e^(delta + 1) <= n )  delta++;
    step = delta / 2;
    d = 0; r = n^(1/2);
    while ( d == 0 && r > 0 ) {
      ord = omega(e,n,r);
      if ( ord > 0 )  d = euclid( e, ord )[0];
      else  r -= step;
    }
    return d;
  }
\end{verbatim}
\end{footnotesize}
The heart of algorithm \emph{rayAttack} is the algorithm 
$\omega(m,n,r)$ determining
an integer $i$ being a multiple of ord$_n(e)$ on the basis of corollary 
\ref{cor-attack-d=e}. Both algorithms use the extended Euclidean 
algorithm \emph{euclid}. In detail:

\begin{footnotesize}
\begin{verbatim}
  /** returns minimum i >= (r - 1) * (n/r - 1) such that m^i = 1 mod n
   *  returns 0 if i is not computable, and -1 if the algorithm fails
   */ 
  long omega( m, n, r ) {
    if ( gcd(m,n) != 1 )  return  0;
    else {
      i = (r - 1) * (n/r - 1);
      m = m % n;
      // determine b such that b * m^i = 1 mod n:
      b = euclid (n, ( m^i % n ) )[1] mod n;
      // determine maximum exponent gamma such that m^gamma divides b:
      gamma = 0;
      for ( k = a.length - 1; k >= 0; k-- ) {
        if ( b >= a[k] ) {
          if ( b % a[k] == 0 ) {
            gamma += 2^k;
            b /= a[k];
          } 
          else break;  // not a power of e
        }
      }
      i += gamma;
      if ( i > 0 &&  b != 1 ) {
        i = - 1;  // algorithm fails!
      }
      return i;
    }
  }
\end{verbatim}
\end{footnotesize}
The classical Euclidean algorithm reads:
\begin{footnotesize}
\begin{verbatim}
  // euclid(m,n) = extended Euclidean algorithm 
  // returning x0, x1 s.t. gcd(m,n) = x0 * m + x1 * n:
  long[] euclid( long m, long n) {
    x[] = {1,0};
    u = 0, v = 1;
    mNegative = false, nNegative = false;
     
    if ( m < 0 ) { m = -m; mNegative = true; }
    if ( n < 0 ) { n = -n; nNegative = true; }
    while ( n > 0 ) {
      // determine q and r such that m = qn + r:
      q = m / n; r = m % n;
      // replace:
      m = n; n = r;
      tmp = u; u = x[0] - q*u; x[0] = tmp; 
      tmp = v; v = x[1] - q*v; x[1] = tmp;
    }
    if  ( mNegative ) x[0] = -x[0];
    if  ( nNegative ) x[1] = -x[1];
    return x;
  }
\end{verbatim}
\end{footnotesize}

\subsubsection{Complexity analysis}
First we note that the running time $T_{\mathrm{euclid}}(m,n)$ of
Euclid's algorithm for two input integers $m,n$ is given by 
\begin{equation}
	\label{T-Euclid}
	T_{\mathrm{euclid}}(m,n) = \log_{\phi}[(3-\phi)\cdot\max(m,n)],
\end{equation}
where $\phi$ is the golden ratio\index{golden ratio} 
$\phi = (1+\sqrt{5})/2$,
see \cite[\S4.5.3, Corollary L (p.360)]{Knuth-1998b}.
If we consider, to simplify, the running time as the number of loops 
to be performed, we therefore we achieve for the running time 
$T_\omega(m,n,r)$ of the $\omega$-function $T_\omega(m,n,r)$ $=$ 
$T_{\mathrm{pow}}(m, \lfloor f(r) \rfloor )$ $+$ 
$T_{\mathrm{euclid}}(n, m^{\lfloor f(r) \rfloor} 
\mbox{ mod } n) + \frac{1}{2}\log_m n + \frac{1}{2}\log_m n$, 
i.e.\ \cite[\S2.12]{Buchmann-2001}
\begin{eqnarray}
\label{T-omega}
	T_\omega(m,n,r) = \log_2 \lfloor f(r) \rfloor \cdot (\log_2 n)^2
	 + \log_\phi [(3-\phi)n] + \log_m n.
\end{eqnarray}
Since the complexity $T_{\mathrm{ray}}(e,pq,r)$ of the ray Attack 
algorithm (with $n$ $=$ $pq$) then is given by 
$$T_{\mathrm{ray}}(e,pq,r) = \frac{r-p}{\log_e pq} T_\omega(e,pq,r)
+ T_{\mathrm{euclid}}(e, \omega(e,pq,r)),$$ 
and since by $\omega(e,pq,r)$ $<$ $n$ we have
$T_{\mathrm{euclid}}(e, \omega(e,pq,r)) < 
T_{\mathrm{euclid}}(e, pq)$, we obtain
\begin{eqnarray}
\label{T-rayAttack}
	T_{\mathrm{ray}} (e,pq,r) &<& 
	  \left( \frac{r-p}{\log_e pq} + 1 \right) 
	   \log_\phi [(3-\phi)\,pq]
	\nonumber \\* & & {}
	  + (r-p) \left( 1 + 
	    \frac{\log_2 \lfloor f(r) \rfloor \cdot (\log_2 pq)^2}
	         {\log_e pq } \right)
	\nonumber \\*
	&=& 
	O\left((r-p) \ln r \cdot \ln e \cdot \ln pq\right).
\end{eqnarray}
(Note that $f(r) = O(r)$.)

\section{Discussion}
In this article a new ansatz to attack RSA cryptosystems is described, 
basing on geometric
properties of the Euler functions, the \emph{Euler function rays}.
However, a resulting algorithm turns out to be inefficient.
It essentially consists of a loop with
starting value determined by
the Euler function ray and with step width given by a function 
$\omega_e(n)$ being a multiple of the order $\mathrm{ord}_n(e)$,
where $e$ denotes the public key exponent and $n$ the RSA modulus.
For $n=pq$ and an estimate $r<\sqrt{pq}$ for the smaller prime 
factor $p$, the running time is given by 
$T(e,n,r) = O((r-p)\ln e \ln n \ln r).$

In other words, this attack is queuing up into a long series of failed attacks
on RSA. So, what is gained in the end? 
First, we achieved a small mathematical novelty,
the Euler function rays, i.e.\ geometrical properties of the Euler
function. 
To my knowledge they have never been mentioned
before. Second, the $\omega$-function has been introduced, being closely
related to the order of a number but being more appropriate for
practical purposes.
Finally, this trial
as another failure in fact is good news. It seems that e-commerce basing on
RSA can go on.

\begin{appendix}
\section{Appendix}
\subsection{Euler's Theorem}
If $n$ is a prime, the set of all numbers (more exactly: of all
residue classes) modulo $n$ is a field with respect to addition and multiplication, as is well known.
However, if $n$ is a composite integer, the ring of all numbers 
modulo $n$ 
is not a field, because the cancellation of a number (more exactly: a congruence) modulo $n$ by any 
divisor $d$ of $n$ also requires the corresponding cancellation of $n$, 
and thus carries us from the ring modulo $n$ to another ring, namely 
modulo $n/d$. In this case, $d$ is said to be a zero divisor of the ring, 
since $d|n$ and $n$ $=$ $n/d$ $=$ 0 $\mbox{mod}$ $n/d$. 
For instance, for $n=9$ the congruence 
$$ 15 = 6 \mbox{ mod } 9$$
is cancelled by $d=3$ through
$${15 \over d} = {6 \over d} \mbox{ mod } {9 \over \mbox{gcd}\,(d,9)},
  \qquad \mbox{or} \qquad 5 = 2 \mbox{ mod } 3.$$
However, if we avoid the zero divisors of $n$ and consider only the those numbers
(more exactly: primitive residue classes) $a \mbox{ mod } n$ with gcd$\,(a,n)=1$, then all divisions by \emph{these} elements can be uniquely performed.
For example, by gcd$\,(5,12)=1$
$$5x = 10 \mbox{ mod } 12
  \qquad \iff \qquad x = 2 \mbox{ mod } 12.$$
These numbers actually constitute a multiplicative group of order $\varphi(n)$:

\begin{defi}
	For $n\in \mathbb{N}$, $n>1$, \emph{Euler's $\varphi$-function} or 
	\emph{totient function} assigns to $n$ the number $\varphi(n)$ of 
	positive integers $k < n$ relatively prime to $n$, i.e.
	\begin{equation}
	 	\varphi(n) = \#\, \mathbb{Z}_{n}^{*}, 
	 	\qquad \mbox{where} \qquad
	 	\mathbb{Z}_{n}^{*} = \{k\in \mathbb{N}: k<n \mbox{ and }
	 	  \mathrm{gcd}\,(k,n) = 1\}.
	 	\label{phi-function}
	 \end{equation} 
\end{defi}
$\mathbb{Z}_{n}^{*}$ is the multiplicative group modulo $n$.
For instance, the set of numbers less than 12 and relatively prime to 12 
are \{1, 5, 7, 11\}, and thus $\varphi(12) = 4$. An explicit formula 
denotes
\begin{equation}
	\varphi(n) = p_{1}^{\alpha_{1}-1} \cdots p_{r}^{\alpha_{r}-1} \cdot
	            (p_{1} -1 ) \cdots ( p_{r}-1)
	           = n \cdot \prod_{p|n} \left( 1 - {1 \over p} \right), 
	\label{phi-eplizit}
\end{equation}
if the prime factorization of $n$ is given by 
$n=p_{1}^{\alpha_{1}} \cdots p_{r}^{\alpha_{r}}$. E.g., $12 = 2^2 \cdot 
3$, and 
$$\varphi(12) = 2 \cdot 2 = 12 \left( 1-\frac12 \right)
                  \left( 1-\frac13 \right)
              = 4.$$
\begin{satz}[Euler's Theorem]
	\label{satz-Euler}
	If $\mathrm{gcd}\,(m,n)=1$, then
	\begin{equation}
		m^{\varphi(n)} = 1 \ \mathrm{mod}\ n.
		\label{Euler-Satz}
	\end{equation} 
\end{satz}
For a proof see, e.g., \cite[\S4.1]{Padberg-1996}.

\subsection{The Carmichael function and Carmichael's Theorem}
Euler's Theorem can be strengthened. As we will see, this will yield an efficient determination of key pairs of a RSA public key cryptosystem, much more efficient than the originally (and yet nowadays in many textbooks) proposed procedure based on Euler's Theorem.

\begin{defi}
	For $n\in \mathbb{N}$ let $n=\prod_{i=1}^r p^{\alpha_{i}}$ be its 
	prime factorisation. Then the 
	\emph{Carmichael\footnote
	{Robert D. Carmichael (1879 -- 1967), U.S.\ mathematician}
	function} $\lambda$ is given by 
	$\lambda(n)$ = $\mathrm{lcm}\,[\lambda(p_{i}^{\alpha_{i}})]_{i}$, where 
	for each $i=1,\ldots, r$,
	\begin{equation}
		  \lambda (p_{i}^{\alpha_{i}}) =
		  \left\{ \begin{array}{ll}
		    2^{\alpha_{i}-2} & \mbox{if } p_{i}=2 \mbox{ and }
		      \alpha_{i} \geqq 3,\\ 
		    p_{i}^{\alpha_{i}-1} (p_{i}-1) & \mbox{otherwise.} 
		  \end{array} \right.
		\label{Carmichael-function}
	\end{equation} 
\end{defi}

For $n>2$, $\lambda(n)$ is even (since $p_{i}-1$ as an even integer 
divides $\lambda(n)$); for $n=2$, we have simply $\lambda(2)$ = 
$\varphi(2)$ = 1. Moreover, since $\lambda(n)$ is the least common 
multiple of factors of $\varphi(n)$, it divides the Euler totient 
function:
\begin{equation}
	2\mid \lambda(n) \mid \varphi(n) \qquad \mbox{for } n>2.
	\label{2-lambda-phi}
\end{equation}

\begin{satz}[Carmichael's Theorem]
	\label{satz-Carmichael}
	If $m,n\in \mathbb{N}$ and $\mathrm{gcd}\,(m,n)=1$, then
	\begin{equation}
		m^{\lambda(n)} = 1 \ \mathrm{mod}\ n.
		\label{Carmichael-Satz}
	\end{equation}
	Moreover, $\lambda(n)$ is the smallest exponent with this property. 
\end{satz}
Using Carmichael's Theorem, we have a way of explicitly writing down the 
quotient of two residue classes $a/b \mbox{ mod }n$. The formula is
\begin{equation}
	{a \over b} = ab^{-1} = ab^{\lambda(n)-1} \ \mathrm{mod}\ n, 
	\qquad \mbox{ if } \mathrm{gcd}\,(b,n) = 1,
	\label{quotient}
\end{equation}
i.e.\ $b^{-1} = b^{\lambda(n)-1} \ \mathrm{mod}\ n$.

\begin{bsp}
	For $n$ $=$ 65\,520 $=$ $2^4$ $\cdot$ $3^2$ $\cdot$ 5 $\cdot$ 7 
	$\cdot$ 13, 	Euler's function assumes the value 
	$\varphi(n)$ $=$ 8 $\cdot$ 6 $\cdot$ 4 $\cdot$ 6 
	$\cdot$ 12 $=$ 13\,824, while $\lambda(n)$ $=$ lcm(4, 6, 4, 6, 12) 
	$=$ 12. 
	For all $m$ with gcd\,($m,n$) $=$ 1 we thus have
	$$ m^{12} = 1 \ \mathrm{mod}\ 65\,520.$$
	For each $m$ with $\mathrm{gcd}\,(b,n)$ $=$ 1
	we have $m^{-1} = m^{11}\ \mathrm{mod}\ 65\,520$. For 
	instance, $$\frac{1}{11} = 11^{11} = 47\,651 \ \mathrm{mod}\ 
	65\,520.$$
\end{bsp}

\begin{satz}
	\label{satz-Carmichael-Fermat}
	If $n\in \mathbb{N}$ is a product of distinct primes, i.e.\
	$n=\prod_{i}p_{i}$, then
	\begin{equation}
		m^{\lambda(n)+1} = m \ \mathrm{mod}\ n
		\qquad \mbox{for all } m\in \mathbb{Z}.
		\label{Carmichael-Fermat}
	\end{equation}
\end{satz}
For a proof see, e.g., \cite[\S{A2}]{Riesel-1994}.

\phantom{p}
If the multiplicative group $\mathbb{Z}_{n}^{*}=\{m : 1 \leqq m, 
\mathrm{gcd}\,(m,n)=1\}$ 
decomposes into the subgroups $G_{i}$, 
\begin{equation}
	\mathbb{Z}_{n}^{*} = G_{1} \times G_{2} \times \ldots \times G_{k},
	\label{M-n-decomposition}
\end{equation} 
and if $d_{i}$ is the order of the group $G_{i}$, then each element 
$m\in \mathbb{Z}_{n}^{*}$ can be written in the form
\begin{equation}
	m = g_{1}^{e_{1}} g_{2}^{e_{2}} \cdots g_{k}^{e_{k}} \qquad \mbox{with }
	    1 \leqq e_{i} \leqq d_{i}.
	\label{m-decomposition}
\end{equation} 
Furthermore, for each $i$,
\begin{equation}
	g_{i}^{d_{i}} = 1 \ \mathrm{mod}\ n, \quad \mbox{with } 
	d_{i}|\lambda(n).
	\label{g-i-decomposition}
\end{equation} 
For instance, $\mathbb{Z}_{15}=\{1,2,4,7,8,11,13,14\}$. We see that 
$\varphi(15) = 8 = \#\, \mathbb{Z}_{15}$. All possible subgroups $G_{i}$ of 
$\mathbb{Z}_{15}$ are the following ones.
$$ G_{1}=\{1\},\ G_{2}=\{1,4\},\ G_{3}=\{1,11\},\ G_{4}=\{1,14\},$$
$$ G_{5}=\{1,2,4,8\},\ G_{6}=\{1,4,7,13\}.$$
Hence $d_{1}=1$, $d_{2}=d_{3}=d_{4}=2$, and $d_{5}=d_{6}=4$. They all 
divide $\lambda(15)=4$.

\begin{cor}
	\label{cor-Carmichael}
	Let be $e,m,n\in \mathbb{N}$, $n>1$, and either $n$ a product of distinct
	primes, or $\mathrm{gcd}\,(m,n)=1$.
	Then for all $e \in \mathbb{N}$
	\begin{equation}
		m^{e} = m^{e\ \mathrm{mod}\ \lambda(n)} \ \mathrm{mod}\ n.
		\label{equ-Carmichael}
	\end{equation} 
\end{cor}
\begin{lem}
	\label{lemma-lambda}
	For $n \in \mathbb{N}$,
	\begin{equation}
		\lambda(n) \leqq n-1.
		\label{lambda-estimate}
	\end{equation}
\end{lem}
\emph{Proof.} Because 
$\lambda(p)<p$ for every prime, $\lambda(n)<n$ as the least common 
multiple of the Carmichael function values of the prime factors of $n$.
\hfill{$\square$}
\end{appendix}



\phantom{pause}

\end{document}